# LEVERAGING FUNDAMENTAL ANALYSIS FOR STOCK TREND PREDICTION FOR PROFIT

**John Phan[1], Hung-Fu Chang[2]**
[1] UIndy School of Business, University of Indianapolis, Indianapolis
[2] R. B. Annis School of Engineering, University of Indianapolis, Indianapolis
phant@uindy.edu, hchang@uindy.edu

**ABSTRACT**

This paper investigates the application of machine learning models, Long Short-Term Memory (LSTM), one-dimensional Convolutional Neural Networks (1D CNN), and Logistic Regression (LR), for predicting stock trends based on fundamental analysis. Unlike most existing studies that predominantly utilize technical or sentiment analysis, we emphasize the use of a company's financial statements and intrinsic value for trend forecasting. Using a dataset of 269 data points from publicly traded companies across various sectors from 2019 to 2023, we employ key financial ratios and the Discounted Cash Flow (DCF) model to formulate two prediction tasks: Annual Stock Price Difference (ASPD) and Difference between Current Stock Price and Intrinsic Value (DCSPIV). These tasks assess the likelihood of annual profit and current profitability, respectively. Our results demonstrate that LR models outperform CNN and LSTM models, achieving an average test accuracy of 74.66% for ASPD and 72.85% for DCSPIV. This study contributes to the limited literature on integrating fundamental analysis into machine learning for stock prediction, offering valuable insights for both academic research and practical investment strategies. By leveraging fundamental data, our approach highlights the potential for long-term stock trend prediction, supporting portfolio managers in their decision-making processes.

## 1   Introduction

The application of artificial intelligence (AI) has significantly expanded across various industries, including finance, prompting a growing interest among researchers in applying machine learning techniques to stock price prediction. Most existing studies in this field use technical analysis or sentiment analysis in machine learning models [1]. However, employing fundamental analysis in machine learning for stock prediction remains limited.

Several reasons lead scientists to use technical analysis and sentiment analysis in machine learning for prediction in the stock market rather than considering fundamental analysis. First, technical analysis is used by short-term traders (e.g., day traders and swing traders) to make trading decisions based on recent stock price movements. For instance, Western Indicators use tools like support and resistance lines, Fibonacci retracements, Simple Moving Averages (SMA), and the Relative Strength Index (RSI) to identify broader market trends and to help traders predict price movements over periods ranging from days to weeks to months, making them suitable for day and swing trading. Aligning the same thoughts inspires the research community to apply technical analysis to various machine learning models for predictions. Second, technical analysis is based on the assumption that stock prices move in trends and that historical patterns will repeat themselves, implying that all publicly available information is already reflected in the stock's price. Third, sentimental analysis leverages textual data, such as news articles, social media posts, and online forums, to gauge investor sentiment toward specific stocks or the broader market. The strategy of following the crowd is the main idea behind sentiment analysis. People believe sentimental analysis reflects collective emotions and psychological behaviors toward the stock market. One can rely on others' research or reaction to market emotions like fear, greed, and herding for decision-making. With the growing amount of data from social media posts, public opinions, and professional reports, analyzing those online data to gauge market sentiment for understanding market behaviors or predicting short-term stock movements has become an attractive approach recently.

Consequently, numerous studies suggested approaches based on sentiment data for estimating potential price movements and market momentum.

In contrast, fundamental analysis is to assess a stock's intrinsic value by examining financial statements and macroeconomic factors disclosed in a company's public reports. A company's intrinsic value is calculated quarterly and annually based on financial reports and future guidance. As a result, fundamental analysis is typically used for medium- to long-term evaluations spanning from several quarters to years, making it a preferred strategy for long-term value investors. Due to this longer-term perspective, fundamental analysis is often excluded from machine learning models for stock prediction, which typically prioritizes shorter time frames. It is perceived as less applicable for short-term stock price prediction, even around one year, because investment returns or target prices based on fundamental analysis are generally expected to be realized over a long period.

This lack of employing fundamental analysis in machine learning models creates a gap in understanding its potential value in stock prediction. For example, researchers who align with Fama's Efficient Market Hypothesis (EMH) [18] and Random Walk Hypothesis view technical analysis, and many others consider even fundamental analysis, as ineffective, advocating that no profitability can be made in an efficient market. However, successful value investors who have consistently beat the S&P 500 have suggested that fundamental analysis, compared to technical analysis, is a valuable tool and has an edge over technical analysis for profit realization. To address this gap, our study explores how AI techniques can be applied to stock price models, offering insights that may benefit portfolio managers employing active or tactical investment strategies. We develop several machine learning models that apply fundamental analysis to predict stock trends based on potential returns. These returns are defined in two ways: the difference in stock prices from the start to the end of the year, as influenced by historical financial data, and the discrepancy between intrinsic values and current stock prices. Our primary contribution in this study is to provide a method that only uses fundamental analysis data in machine learning models. This approach enhances the understanding of how fundamental analysis alone can be applied in stock trend predictions, contributing new insights for both academic research and practical applications in the financial industry.

## 2   Literature Review

Past research regarding machine learning on stock predictions can be classified into three main categories according to the analysis techniques used in the model. We describe them in the following three sections.

### 2.1   Technical analysis

Aadhitya et al. developed a CNN-LSTM neural network model to analyze daily stock prices of companies listed on the NIFTY-50, NYSE, and NASDAQ from 2000 to 2021. Their model achieved high accuracy, particularly for NIFTY stocks, reaching up to 99%, although accuracy varied for NYSE and NASDAQ stocks. Compared to other models, such as standalone LSTM and XGBoost, their model demonstrated superior performance with minimal error and variance [8]. Similarly, Nelson et al. applied an LSTM model to predict stock prices on the Brazilian stock exchange, achieving a 55% accuracy rate. They emphasized that technical analysis relies on patterns in stock prices driven by supply and demand, which tend to repeat but do not account for external factors such as political or economic events. Their findings suggest that while the LSTM model offers higher accuracy, reducing variance is crucial for its reliability in stock prediction [9]. These studies indicate that while machine learning models like CNN-LSTM and LSTM enhance predictive capabilities for short-term forecasting, they require careful model selection and optimization due to their inherent limitations.

### 2.2   Sentimental analysis

Sentiment analysis evaluates text data to predict short-term stock price movements by identifying positive or negative sentiment. Ding et al. [6] found that financial news events were better predictors of stock prices than simple word counts, with deep neural networks outperforming linear models by learning hidden relationships between events and stock prices. However, they noted that combining company news with sector news reduced prediction accuracy due to noise from irrelevant information. Zhang et al. [10] further explored sentiment analysis using financial statements, earnings scripts, and social media, finding that while public sentiment can provide insights, sentiments from professional sources are often biased and less reliable. Despite these challenges, machine learning models trained on sentiment data have shown potential in predicting stock market trends by integrating structured financial data with unstructured textual information.

### 2.3 Fundamental analysis

Recently, few studies started to use fundamental analysis in machine learning for stock prediction or financial forecasting. Huang et al. [2] combined historical financial data of large-cap stocks from the S&P 100 index with models like Feed-forward Neural Networks (FNN), Random Forest (RF), and Adaptive Neural Fuzzy Inference System (ANFIS) to predict long-term stock performance prediction. Their experimental results showed that all three methods are capable of constructing stock portfolios that outperform the market without any input of expert knowledge if they are fed with enough data in which Random Forest showed the best result. [12] Bekiros and Georgoutsos' research shows that, without trading costs, the return of the neuro-fuzzy model consistently outperforms the recurrent neural model, as well as the buy and hold strategy during bear period. Whereas, during the bull period, the buy and hold strategy produces higher returns than neuro-fuzzy models or neural networks. Ftiakas et al. [3] applied seven different algorithms to 1,353 NASDAQ stocks, showing that no single algorithm is universally superior, underscoring the need for multiple approaches in financial analysis. Cao and You [4] analyzed historical data from 1965 to 2019, finding that approaches like Random Forest, Gradient Boosting, and Artificial Neural Networks provided more accurate forecasts than traditional methods by uncovering subtle nonlinear relationships in financial data. These findings highlight the potential of machine learning to refine traditional fundamental analysis and provide new insights for investment decisions.

### 2.4 Summary

In summary, while traditional fundamental analysis focuses on evaluating a company's financial health through detailed analysis of financial statements, the integration of machine learning techniques can enhance the accuracy and depth of these evaluations. This combined approach provides a more comprehensive understanding of a company's value, considering both quantitative and qualitative factors, and helps investors stay ahead in an increasingly complex financial landscape.

## 3 Method

The purpose of this paper is to explore the connection between machine learning techniques and fundamental analysis, thus, we gathered the companies' fundamental information between three financial statements. Using the raw numbers from all three statements, we also calculated the key financial ratios such as liquidity ratios and profit margins individually, as well as indices' averages.

### 3.1 Data Collection

The fundamental data was retrieved from Yahoo Finance in the range of 5 years from the first day of 2019 to the last day of 2023. Each selected company is expected to have 5 records. Unfortunately, some might miss one entry because they do not have all 5 years' worth of stock prices. From the above criteria, we resulted in picking stocks from very stable indexes including the Industrial Sector (XLI), Utility Sector (XLU), Consumer Staple Sector (XLP), Consumer Discretionary Sector (XLY), Dow Jones Index (DJI), and Top 100 companies in the U.S. stock exchange (QQQ). Since there are a lot of overlaps between the index, we eliminate the overlapping companies and come to the final count of 269 publicly traded companies.

One strategy that was used in past research [10] for getting companies for the machine learning dataset for stock performance prediction proved to avoid random selection or high volatility. This is also the method aligning with what industrial professionals believe about the usage of fundamental analysis. Often, companies with high volatility are small and their stock price can be highly impacted by the sentiment of the market (e.g., irrational optimism from the majority of retail traders in the market).

To avoid unwanted influence on price movement, we chose to follow the common belief that the bigger the market capitalization, the less volatile the company is due to a higher dollar per share, lower revenue growth expectation, more consistency in their revenue guidance, and the risks are more manageable due to their economy of scales and resources. Moreover, we also believe that the biggest companies within the "stable" or consumer goods sectors are less volatile than high growth expectation technology companies due to their needs for our daily consumption. Thus, in summary, we decided to pick companies that show the most consistency in their price movement and replicate the market as closely as possible; at the same time, by avoiding the influence of high volatility, we can better understand the connection between fundamental analysis, the price movement and machine learning algorithms.

## 3.2 Dataset

Our training data is sourced from three key financial documents: the income statement, balance sheet, and cash flow statement, which were retrieved from Yahoo Finance for the period spanning 2019 to 2023. These documents provide comprehensive historical financial data for the selected companies, offering a detailed view of their financial performance and position over time. From this data, we determine features and labels that serve as the base for our machine learning models.

### 3.2.1 Features

The machine learning model's features consist of a company's raw historical financial data and various financial ratios calculated from the income statement, balance sheet, and cash flow statement (see Table 1). The model also includes each company's intrinsic value, determined using the Discounted Cash Flow (DCF) model.

Table 1: Features in Machine Learning Models

| **Historical Financial Data** | **Income Statement:** |
| --- | --- |
| | Total Revenue, Cost of Revenue, SG&A, R&D, Operating Expenses, Net Income, Diluted EPS, Diluted Average Shares, Net Interest Income, EBITDA, EBIT |
| | **Balance Sheet:** |
| | Long Term Debt, Total Debt, Invested Capital, Working Capital, Stockholders Equity, Retained Earnings, Total Asset, Cash & Cash Equiv, Inventory, Gross PPE, Current Assets, Current Liabilities, Total Liabilities |
| | **Cash Flow Statement:** |
| | Net Income, Depreciation & Amortization, Gain/Loss on Business Sale, Impairment Charge, Change in Working Cap, Operating Cash Flow, Net PPE and Sale, Net Tangible Purchase and Sale, Net Business Purchase and Sale, Net Investment Purchase and Sale, Investing Cash Flow, Net Common Stock Issuance, Repurchase of Capital Stock, Cash Dividends Paid, Financing Cash Flow, Change in Cash, Capital Expenditures, Issuance of Debt, Repayment of Debt, Free Cash Flow |
| **Financial Ratio** | Current Ratio, Cash Ratio, Quick Ratio, Debt to Asset Ratio, Debt to Equity Ratio, Gross Margin, Operating Margin, EBITDA Margin, Net Margin, Interest Coverage Ratio, Free Cash Flow Margin |
| **Discounted Cash Flow Model Attributes** | Growth Rate1, Growth Rate2, Growth Rate3, Forecasted Revenue1, Forecasted Revenue2, Forecasted Revenue3, Forecasted EBITDA1, Forecasted EBITDA2, Forecasted EBITDA3, EV / EBITDA Multiple, Beta |
| | Risk Free 10-Year Treasury Rate, Market S&P 500 10-Year Return, Corporate Tax Rate (21%), Intrinsic Value1, Intrinsic Value2, Intrinsic Value3, Final Intrinsic Value |

The rationale behind the data selection for features is that fundamental analysis is mostly used for finding out how much the financial numbers and ratios would affect the stock price of a company. In practice, looking at the numbers and knowing the ratios is only part of the job of an investment analyst. These ratios are important numbers that help investors compare each company to their competitors and price them accordingly. Moreover, the financial ratios we calculated are based on the DCF model which is the most commonly used technique to value a company. Using the DCF model aligns with investment practices that determine how much they are willing to pay for a company based on its projected cash flows over the next 5 to 10 years. As a result, Eq. (1) to (11) describes all the calculated attributes that we used in the financial ratio part.

$$Current\ Ratio\ =\frac{Current\ Assets}{Current\ Liabilities} \tag{1}$$

$$Cash\ Ratio\ =\frac{Cash\ and\ Cash\ Equiv}{Current\ Liabilities} \tag{2}$$

$$Quick\ Ratio\ =\frac{(Current\ Assets-Inventory)}{Current\ Liabilities} \tag{3}$$

$$Debt\ to\ Asset\ Ratio\ =\frac{Total\ Debt}{Total\ Assets} \tag{4}$$

$$Debt\ to\ Equity\ Ratio\ =\frac{Total\ Debt}{Stockholders\ Equity} \tag{5}$$

$$Gross\ Margin\ =\frac{Gross\ Profit}{Total\ Revenue} \tag{6}$$

$$Operating\ Margin\ =\frac{Operating\ Income}{Total\ Revenue} \tag{7}$$

$$EBITDA\ Margin\ =\frac{EBITDA}{Total\ Revenue} \tag{8}$$

$$Net\ Margin\ =\frac{Net\ Income}{Total\ Revenue} \tag{9}$$

$$Interest\ Coverage\ Ratio\ =\frac{EBIT}{Net\ Interest\ Income} \tag{10}$$

$$Free\ Cash\ Flow\ Margin\ =\frac{Free\ Cash\ Flow}{Total\ Revenue} \tag{11}$$

### 3.2.2 Labels

We would like to use machine learning models to perform two predictions on a company: (1) Annual Stock Price Difference (ASPD) and (2) Difference between Current Stock Price and Intrinsic Value (DCSPIV). These two labels both indicate profit. On one hand, the investor can decide to sell a stock for profit anytime in a year and the price changes every second according to what is available to the public, financially. Specifically, when a company's stock price at the start of the year is positively affected by the financial announcement or guidance, increasing to a higher point at year's end, the label is 1; otherwise, it is 0. On the other hand, the other straightforward strategy for profit is to use intrinsic value to determine whether to purchase or sell a stock at this very moment because the intrinsic value is used to assess whether a company is trading at a discount or premium in the current market state. In this scenario, the difference between current stock price and intrinsic value is the potential profit or loss for that investment. If that investment results in a gain, the label is 1; otherwise, it is 0.

Although various methods exist to determine a company's intrinsic value across sectors, we aimed to create a formula that is both universally applicable and aligned with investment industry standards. We utilized the

Discounted Cash Flow (DCF) method, incorporating EV/EBITDA multiples, and focused on growth rates, which are the most sensitive factor in the DCF model. We used three growth rates: the historical 5-year rate, the industry or index average, and the Yahoo Finance growth rate. By averaging these intrinsic values and comparing them to the current price, we derive the label.

We labeled the same datasets in two different ways. In the first method, if historical financial data positively impacted the company's stock price over the year, it was labeled as 1; if it had a negative effect, it was labeled as 0. For the second method, we tested the efficiency of LSTM and CNN models using raw and calculated financial ratios along with intrinsic values to predict whether a company's intrinsic value would exceed its current stock price (labeled as 1) or not (labeled as 0).

The label for annual stock price difference (*ASPD*) is defined as the following:

$$ASPD = \{1 \text{ if } P_{ae} \geq P_{ab},\ 0 \text{ if } P_{ae} < P_{ab}\} \tag{12}$$

where, $P_{ae}$ is the stock price on the beginning day of the year and $P_{ae}$ is the stock price on the last day of the year. We also create another label for the difference between stock price and intrinsic value (i.e., DCSPIV). In Eq. (13), $P_{cur}$ is the current stock price (i.e., daily closing stock price), and *I* is the intrinsic value of a company.

$$DCSPIV = \{1 \text{ if } I \geq P_{cur},\ 0 \text{ if } I < P_{cur}\} \tag{13}$$

The current stock price $P_{cur}$ can be viewed as the dynamic value each day that the market thinks the company should be worth. In Eq.(13), what we consider is the company's value changes. We do not consider using stock price difference, for example, the current price $P_{cur}$ subtracts the price at the beginning of the year $\boldsymbol{P_{ab}}$ (i.e., $P_{cur}$ - $\boldsymbol{P_{ab}}$) because $\boldsymbol{P_{ab}}$ can be impacted by temporary market sentiment or information. Using stock price as a comparison baseline can also introduce noises when our study focuses on the value changes.

### 3.3 Machine Learning Models

The LSTM is frequently utilized in stock price prediction research due to its effectiveness in modeling sequences and time-series data, which aligns well with the nature of historical stock prices. An LSTM is composed of a series of interconnected memory cells, each containing three critical components: the Input Gate, Forget Gate, and Output Gate. These gates enable the LSTM to retain and update relevant information over extended sequences, facilitating the transmission of knowledge across the network nodes.

While Convolutional Neural Network (CNN) are traditionally used in image processing, they are also highly effective in identifying patterns, trends, and anomalies within datasets—tasks essential for decision-making and risk management in finance. In the financial sector, CNN is applied for earnings forecasting, anomaly detection in credit transactions, and recognizing various market conditions or shifts. Unlike LSTM networks, which are designed to capture long-term dependencies in data sequences, one-dimensional CNN (1D CNN) are particularly advantageous for detecting short-term, abrupt trends or changes in the data.

The Logistic Regression (LR) is simple. It performs well for linearly separable data and is highly interpretable, making it useful in fields like medicine, finance, and social sciences. It also serves as a foundation for more complex models, such as neural networks. Therefore, logistic regression is widely used for binary classification tasks, where the outcome variable is categorical with two possible values (e.g., 0 and 1). Its binary outcome is calculated by the logistic function (or sigmoid function), which maps any real-valued number into a value between 0 and 1, representing the probability. In logistic regression, the model estimates the coefficients of the input variables to best fit the data. Once the model is trained, it predicts the probability that a new input belongs to a specific class. If the probability exceeds a threshold (commonly 0.5), the model assigns the input to one class; otherwise, it assigns it to the other.

According to different labels, we construct the same type of models slightly differently. The following describes the details of a model architecture.

- **LSTM on *ASPD*:**

This Long Short-Term Memory (LSTM) neural network was specifically built for binary classification, designed to capture temporal dependencies in stock price data. The preprocessing involved selecting key features, scaling them with StandardScaler, and splitting the dataset into training and testing sets. The architecture consisted of an LSTM layer with two hidden layers, followed by a linear layer, which condensed the sequence output into a single binary prediction via a sigmoid function. The model was trained using Binary Cross-Entropy with Logits Loss and the Adam optimizer for 5,000 epochs. During training, the model tracked performance metrics such as loss, accuracy, precision, recall, and F1 score to evaluate its learning ability.

- **CNN on *ASPD*:**

Our CNN architecture consisted of two Conv1D layers, each followed by a max pooling layer to reduce dimensionality and extract important features from the sequence. After the convolutional layers, the data was flattened and passed through two fully connected layers, with the final layer providing the binary classification output using a sigmoid activation function.

- **LSTM on *DCSPIV*:**

The model was compiled with Binary Cross-Entropy for the classification task and Mean Squared Error (MSE) for the regression task. The preprocessing involved filling missing values, flattening arrays, and scaling features using StandardScaler before reshaping the data for sequential input. The model architecture featured a shared LSTM layer with 50 units to process the time-series data, followed by a dense layer with 64 units using ReLU activation. The model had two outputs: one for binary classification (predicting a label) using a sigmoid activation and one for regression (predicting the intrinsic value) using a linear activation.

- **CNN for *DCSPIV*:**

The final model used a CNN for the same multi-output task on the intrinsic value dataset. Data preprocessing included filling in missing values, flattening arrays, and scaling the features before reshaping them to a format compatible with the CNN architecture. The CNN had two convolutional layers, each followed by max pooling to reduce dimensionality and extract key patterns. After flattening the output, the data was passed through a dense layer before branching into two outputs: one for binary classification (with a sigmoid activation) and another for regression (with a linear activation).

## 4 Results

After running 5,000 epochs for each model for ASPD and DCSPIV, the CNN model performed the worst among all of them, achieving just 55.36% accuracy on ASPD (see Table 2). Whereas Logistics Regression achieved the best average testing accuracy out of three models for both datasets at 74.66% accuracy in *ASPD* and 72.85% in DCSPIV.

We notice that LSTM and CNN both perform well in the training, having average accuracy of 98.51% and 97.51%, respectively, which outperform Logistic Regression. However, their testing accuracy is poorer than Logistic Regression. Among all the LSTM values in both ASPD and DCSPIV, LSTM's average recall value reaches 92.66% in DCSPIV. That means LSTM can detect true positive cases well in DCSPIV. Overall, regarding average testing recall, F1 score, precision, and accuracy, we think the performance of CNN is not as good as the other two models because only CNN's average testing precision (i.e., 58.35%) in DCSPIV is slightly better than LSTM. The Logistics Regression performs best in terms of average testing accuracy, precision, recall, and F1 score.

Table 2: Machine Learning Results

|  | ASPD | | | DCSPIV | | |
| --- | --- | --- | --- | --- | --- | --- |
|  | LSTM | CNN | LR | LSTM | CNN | LR |
| **Average Training Accuracy** | 0.9851 | 0.9751 | 0.7950 | 0.7174 | 0.7331 | 0.7937 |
| **Average Train Precision** | 0.9855 | 0.9765 | 0.7770 | 0.7140 | 0.7475 | 0.7533 |
| **Average Training Recall** | 0.9950 | 0.9876 | 0.9432 | 0.9305 | 0.8919 | 0.9530 |
| **Average Training F1-Score** | 0.9895 | 0.9814 | 0.8519 | 0.8073 | 0.8109 | 0.8410 |
| **Average Test Accuracy** | 0.5763 | 0.5536 | 0.7466 | 0.5958 | 0.5899 | 0.7285 |
| **Average Test Precision** | 0.6243 | 0.6102 | 0.7571 | 0.5784 | 0.5835 | 0.6961 |
| **Average Test Recall** | 0.7169 | 0.6834 | 0.8962 | 0.9266 | 0.8351 | 0.9346 |
| **Average Test F1-Score** | 0.6669 | 0.6444 | 0.8124 | 0.7116 | 0.6856 | 0.7975 |

## 5 Discussion

We discuss our outcomes in three different perspectives, model performance, approach comparison, and limitation.

### 5.1 Model Performance

Logistic Regression is regarded as an effective approach when the data is not sequential or spatial in nature. LSTM is more appropriate for sequential data and CNN can exact spatial features. Because Logistic Regression provides the best performance, this can imply that features extracted from fundamental analysis might not equip sequential or spatial characteristics. In addition, we think Logistic Regression demonstrates robust generalization on both ASPD and DCSPIV datasets despite having lower training metrics. While LSTM and CNN have strong training performances, indicating their capacity to learn complex patterns, they tend to overfit, leading to a noticeable drop in test accuracy, precision, and F1-score. LR, being a simpler model, achieves higher accuracy, precision, and F1-score on the test sets for both datasets, suggesting that it is less sensitive to overfitting and better at generalizing across different data types. Though LSTM maintains a consistently high recall, indicating a strong ability to detect true positives, LR's balanced performance between precision and recall on the test data makes it a more effective and stable choice for these datasets.

LSTM outperforms CNN in every aspect in ASPD. This does align with our original estimation since LSTM is often better for predicting the time series data. It seems that this architecture struggled to generalize the stock price differences over time, potentially due to the complexity or noisiness of the data. The LSTM model on this DCSPIV is significantly better than on ASPD. This suggests that the LSTM's ability to capture temporal dependencies was particularly useful for handling the intrinsic value prediction, which likely involved more sequential complexity and nuance than the stock price data. This CNN on DCSPIV effectively handled both classification and regression tasks, leveraging shared convolutional layers for feature extraction. However, it didn't quite match the LSTM's performance, likely because the temporal patterns in the intrinsic value dataset were better captured by the recurrent nature of the LSTM.

### 5.2 Approach Comparison

Komori's research [16], which supports fundamentalists, utilized a Convolutional Neural Network (CNN) to analyze 2D candlestick charts of the S&P 500 index from 1985 to 2020. Using a simple moving average as a technical indicator, they tested a CNN model, Inception v3, for 1, 3, and 5-day forecasts. Despite CNN's high 78.1% accuracy on the ImageNet dataset, its highest stock market forecasting accuracy was only 50% at 3 days, closely aligning with the Random Walk Hypothesis. This hypothesis argues that stock prices are essentially unpredictable, akin to random coin flips, providing a 50% chance of success over time. Nelson, Pereira, and Oliveira [17] took a different route by applying an LSTM model to 15-minute interval price data from the Brazilian stock exchange, achieving a modest improvement with 55.9% accuracy, a precision of 56.3%, recall of 35%, and an F1 score of 43.1%. Despite using machine learning, both studies barely surpassed the 50% accuracy, no greater than a coin flip.

Our approach, however, achieved higher accuracy, ranging from 55% to 75% in predicting both ASPD and DCSPIV using LSTM, CNN, and Logistic Regression models. Instead of predicting short-term price movements like the previous studies, our results imply that incorporating stock trends with machine learning models can better capture market behavior. Furthermore, our models provide investors with a more profitable strategy by selecting stocks based on traditional fundamental analysis techniques. Although our accuracy only slightly surpasses Komori's findings, it aligns with Nelson's results, though their focus on short-term forecasts contrasts with our longer time horizon. This crucial difference offers traders and investors a better chance to anticipate market directions and manage risks during uncertainty.

Additionally, we applied a basic Discounted Cash Flow (DCF) with an EBITDA for intrinsic value calculation, achieving an accuracy range between 60% and 73%. This outcome suggests that even a simple fundamental analysis model, when applied to machine learning techniques, can produce results comparable to the widely used technical analysis methods. However, we believe more sophisticated fundamental models, tailored to specific sectors, could enhance accuracy even further.

## 5.3 Limitation

Although all our models can reach more than 55% accuracy, this study has several limitations that could be addressed to improve the models' utility for investors. One key limitation is the dataset's size and scope. The time range of data collection was limited, potentially hindering the models' ability to capture long-term trends and fluctuations in the market. Expanding the dataset to cover a broader time span would allow the models to better generalize across various market conditions. Additionally, the number of stocks chosen for analysis was relatively small, which may limit the models' applicability to different sectors and market environments. Including a more diverse set of stocks would improve the robustness of the models across industries. Moreover, financial ratios and key attributes differ in importance depending on the sector. Customizing the model parameters for specific industries could significantly enhance accuracy and relevance, making the models more effective for practitioners. Addressing these limitations through larger datasets, longer timeframes, and industry-specific adjustments could lead to more practical and reliable investment strategies.

## 6 Conclusion

Our study explored the use of LSTM, CNN, and logistic regression (LR) models in integrating fundamental analysis for predicting stock trends, achieving accuracy rates close to 72%. Our findings indicate that all models are better suited for fundamental analysis than technical analysis. Our method can offer a straightforward approach to identifying profitable stocks. By leveraging financial statements and intrinsic value calculations, investors can enhance their decision-making processes, particularly by considering early-year stock purchases. Our research provides valuable insights for both industry professionals and academic researchers, highlighting the potential of machine learning models, which result in long-term, profitable investment strategies.